\documentstyle[twocolumn,prb,aps,psfig,floats]{revtex} 
\begin{document}
\twocolumn[\hsize\textwidth\columnwidth\hsize\csname 
@twocolumnfalse\endcsname
\title{Gauge factor enhancement driven by heterogeneity in
thick-film resistors} 
\author{C. Grimaldi, P. Ryser, and S. Str\"assler} 
\address{D\'epartement de Microtechnique, IPM,
\'Ecole Polytechnique F\'ed\'erale de Lausanne,
CH-1015 Lausanne, Switzerland.}
\maketitle

\centerline \\

\begin{abstract}
We present a simple picture of the gauge factor (GF) enhancement in
highly heterogeneous materials such as thick-film resistors.
We show that when the conducting phase is stiffer than the insulating
one, the local strains within this latter are enhanced with respect
to the averaged macroscopic strain. Within a simple model
of electron tunneling processes, we show that the enhanced local
strain leads to values of GF higher than those expected for a
homogeneous system. Moreover we provide formulas relating
the enhancement of GF to the elastic and microstructural 
characteristics of TFRs.

PACS number(s) 72.20.Fr, 72.80.Tm, 62.20.Dc
\end{abstract}

\vskip 2pc ] 

\narrowtext
\centerline \\


\section{introduction}
\label{intro} 
Thick-film resistors (TFRs) are successfully used as piezoresistive
sensors due to their high strain-sensitivity resistance. 
Their gauge factors (GFs), conventionally defined as
${\rm GF}=\delta R/R\epsilon$, where $\delta R$ is the change of resistance
$R$ for an applied strain $\epsilon$, range in fact from ${\rm GF}\sim 2$ up to
${\rm GF}\sim 35$ at room temperature.\cite{prude1}
Of course, higher values of GF lead to a better piezoresistive response of 
a TFR, so that the knowledge of the factors which enhance GF is of
primary importance in improving piezoresistive sensor performances.

Among these factors, the microstructure of TFRs plays certainly
an important role in enhancing GF. A typical TFR is a composite material
in which metallic grains are embedded in an insulating glassy matrix, and
a positive correlation between the mean metallic grain size and GF has been
established already since some time.\cite{carcia,prude2,prude3}
Fabrication processes such as the peak firing temperature $T_{\rm f}$
modify the compositional and microstructure properties leading to
important and complex modifications of GF. Typically, GF initially
increases with $T_{\rm f}$ and after having reached a maximum, it
shows a more or less pronounced decrease.\cite{prude1} 
Other important intrinsic factors affecting GF are the chemical compositions 
of both the metallic and insulating phases.
Finally, in addition to these effects, there exists an empirical 
relationship between (sheet) resistance $R$ and GF:
\begin{equation}
\label{GF1}
{\rm GF} \propto \ln R,
\end{equation} 
which has been established for room temperatures. According to this
relation therefore the most resistive TFRs show also the higher GFs.

Although these phenomenological properties are useful in improving 
the strain sensitivity in TFRs, the microscopic physical mechanisms
enhancing GF are unclear, and the quest for high GF values rests only
on empirical basis. An even more annoying aspect of this situation
is that simple tunneling-like formulas for $R$ are completely inadequate
to explain the observed GF values. For example, at sufficiently high
temperatures, transport is thought to be governed by tunneling between 
adjacent metallic grains separated by a mean distance $d$:\cite{note1}
\begin{equation}
\label{R1}
R\simeq R_0\exp({2d/\xi}),
\end{equation}
where $\xi=\hbar/\sqrt{2mV}$ is the localization length and
$m$ and $V$ are the electron mass and the tunneling barrier potential, 
respectively. The prefactor in Eq.(\ref{R1}) is $R_0=\Phi R_{\rm H}$
where $R_{\rm H}=2\pi\hbar/e^2\simeq 26$ K$\Omega$ is the Hall resistance
and $\Phi$ is a dimensionless geometrical factor. Assuming homogeneity,
the main dependence on an applied strain $\epsilon$ comes from the
exponent of Eq.(\ref{R1}) through $d\rightarrow d(1+\epsilon)$, and
the gauge factor reduces to:
\begin{equation}
\label{GF2}
{\rm GF}\equiv \frac{\delta R}{\epsilon R}\simeq 2d/\xi.
\end{equation} 
Equations (\ref{R1}) and (\ref{GF2}) are fully consistent with the
empirical relation in Eq.(\ref{GF1}), and $R$ can be rewritten as
$R\simeq R_0 \exp({\rm GF})$. However, typical values of the sheet resistance are
of order $10-1000$ K$\Omega$, while for example for ${\rm GF}=20$ we get
$R\simeq R_0\exp20\sim 10^{10}$ K$\Omega$!
This astronomically large value of $R$ obtained by the simple tunneling argument 
gives an idea of the problem concerning the physical mechanism of GF in TFRs.

In this paper we present a theory of GF in TFRs which, on one hand,
reconciles the observed values of GF with those of sheet resistances and,
on the other hand, clarifies the key elements responsible for 
high gauge factor values. Basically, we claim that the high GF values are 
due to pronounced compositional heterogeneity of TFRs, so that the local strain
$\epsilon_{\rm loc}$ can be substantially different from the applied 
macroscopic strain $\epsilon$.
In this situation, the mean inter-grain distance is modified by the local
strain, $d\rightarrow d(1+\epsilon_{\rm loc})$, so that Eq.(\ref{GF2})
acquires a prefactor:
\begin{equation}
\label{GF3}
{\rm GF}\simeq(2d/\xi)\frac{\epsilon_{\rm loc}}{\epsilon}.
\end{equation}
The value of $\epsilon_{\rm loc}/\epsilon$ depends on the elastic properties of
the metallic and glassy phases as well as on the mean grain size and
inter-grain distance. We show that when the insulating phase is less stiff
than the conducting one, $\epsilon_{\rm loc}$ can be
substantially larger than $\epsilon$ leading therefore to an enhancement
of GF.
In the next section we provide analytical formulas relating
$\epsilon_{\rm loc}/\epsilon$ to the compositional and microstructural
characteristics of TFRs by using a simplified model of electron hopping and
strain distribution.

\section{Origin and basic effects of strain heterogeneity}
\label{origin}

Heterogeneity is an intrinsic feature of TFRs. As already stated in the 
introduction, TFRs are two-phases materials where metallic particles of 
sizes ranging between $\sim 200$ and $\sim 6000$ \AA  are embedded into
an insulating glassy material. Depending on the fabrication processes, 
a certain amount of conducting material is dispersed in the glass and recent x-ray
investigations have revealed that much smaller metallic particles
(with sizes of order $\sim 20$-$70$ \AA) are actually dispersed in the
glass.\cite{mene}
At the present, it is not clear whether the electrons tunnel mainly 
between the large or the small particles, or if possible impurity states 
in the glass play a positive role in the observed conductivities.
Several models of transport in TFRs have been 
proposed,\cite{pike,sheng,prude4,schoepe,roman,flachbart} having however 
the tunneling process as the common element. Hence, in the forthcoming 
discussion we assume Eq.(\ref{R1}) to capture the essential physics at 
least at high temperatures, and interpret $d$ as being a typical tunneling 
distance ignoring for the moment whether it refers to the large or the small
grain mean separation.

An important cause of heterogeneity is the chemical composition of both
the metallic and insulating phases. The empirical search for high GF
values has selected some transition-metal oxides containing heavy elements
(RuO$_2$, IrO$_2$, Bi$_2$Ru$_2$O$_7$ are typical examples) to be
the metallic constituents which present the best TFR performances.
An interesting aspect of this selection is that such metal-oxides are hard
materials characterized by high values of bulk moduli $B$. For example
RuO$_2$, which has $B\simeq 270$ GPa,\cite{hazen} is among the
stiffest materials and IrO$_2$ should have comparable values of $B$.\cite{note2}
Conversely, the glassy phase has $B$ values 
typically of order $40$-$80$ GPa depending on composition. 
Hence, the metallic phase is much 
stiffer than the insulating one so that there is a pronounced 
microscopical heterogeneity in the elastic properties of TFRs.

A straightforward consequence of this elastic heterogeneity is that
TFRs under an applied macroscopic strain $\epsilon$ should
develop highly variable local strain fields. In this case, Eq.(\ref{GF1})
is inadequate because it has been obtained by assuming that the 
microscopic tunneling process, Eq.(\ref{R1}), is affected only by the
macroscopic strain. Hence, a correct treatment of the GF problem
in TFRs would require the knowledge of the local strains developed in
a disordered two-phases material with different elastic properties.
This is a quite difficult task which could be attacked only by
extensive numerical calculations. However this is outside the
scope of this paper, instead our aim here it is to show that
when a system is composed by interconnecting stiff conducting and 
soft insulating phases, the local strain within this latter is 
enhanced compared to the averaged one, and the tunneling process
gets a higher strain sensitivity. 

To illustrate in the simplest way the local strain enhancement
effect, we consider a one-dimensional model
defined as a periodic arrangement of $N$ conducting segments
each of length $s$ separated each other by $N$ insulating 
segments of length $d$, so that the total
length of the one-dimensional array is $L=N(s+d)$. 
The stress-strain relation is simply $\sigma(x)=2\mu(x)\epsilon(x)$,
where $\sigma(x)$ and $\epsilon(x)$ are the stress and strain functions
which depend on the position $x$ along the array. In the two-phases model here
considered, the elastic function $\mu(x)$ assumes the values $\mu_0$ or
$\mu_1$ depending whether $x$ lies within the insulating or the conducting
segments. The equilibrium strain distribution is governed by the equation
$d\sigma(x)/dx=0$ from which it is found that $\epsilon(x)$ is constant
within each segment assuming the value $\epsilon(x)=\epsilon^0$ 
($\epsilon(x)=\epsilon^1$) for $x$ within the insulating (conducting) 
segments, where $\mu_0\epsilon^0=\mu_1\epsilon^1$.

The local values $\epsilon^0$ and $\epsilon^1$ are related to the
macroscopic strain $\epsilon$ by the following relation:
\begin{equation}
\label{strain1}
\epsilon=\frac{1}{L}\int_0^L\! dx \,\epsilon(x)=
\frac{d\epsilon^0+s\epsilon^1}{d+s},
\end{equation}
from which we find that the strain within the insulating segments is
\begin{equation}
\label{strain2}
\epsilon^0=\epsilon\frac{1+s/d}{1+(s/d)(\mu_0/\mu_1)},
\end{equation}
while within the conducting segments is
\begin{equation}
\label{strain3}
\epsilon^1=\epsilon\frac{1+s/d}{1+(s/d)(\mu_0/\mu_1)}\frac{\mu_0}{\mu_1}.
\end{equation}
Equations (\ref{strain2}) and (\ref{strain3}) merely illustrate that,
if $\mu_0\neq \mu_1$, the strain is concentrated more within the
softer phase. For example, in the limiting case where the conducting segments
are perfectly rigid ($\mu_1 \rightarrow \infty$) we find $\epsilon^1=0$
and $\epsilon^0=\epsilon(1+s/d)$.

To investigate the effect of this heterogeneous strain distribution
on transport, we model the electron tunneling 
between two adjacent conducting segments with a tunneling
probability proportional to $\exp(2d/\xi)$, so that, if we neglect the
contribution from the metallic segments, the total resistance reduces to:
\begin{equation}
\label{R2}
R=N R_0\exp(2d/\xi) .
\end{equation}
An applied macroscopic strain $\epsilon$ modifies the tunneling distance
$d$ through the local strain, $d\rightarrow d(1+\epsilon^0)$, so that
the gauge factor ${\rm GF}=\delta R/\epsilon R$ is
\begin{equation}
\label{GF4}
{\rm GF}=(2d/\xi)\frac{\epsilon^0}{\epsilon}.
\end{equation}
By substituting Eq.(\ref{strain2}) in Eq.(\ref{GF4}), we then
obtain
\begin{equation}
\label{GF5}
{\rm GF}=(2d/\xi)\frac{1+s/d}{1+(s/d)(\mu_0/\mu_1)}.
\end{equation}
The above expression clearly illustrate the enhancement effect of heterogeneity
driven by both geometrical, $s/d$, and elastic, $\mu_0/\mu_1$,
parameters. In fact, for metallic phases stiffer than the insulating one,
$\mu_1 > \mu_0$, GF is always larger than $2d/\xi$, reaching the limiting
value
\begin{equation}
\label{GF6}
{\rm GF}=(2d/\xi)\frac{\mu_1}{\mu_0},
\end{equation}
in case of tunneling between large conducting phases ($s/d\gg 1$). 
Another limiting case of Eq.(\ref{GF5}) is given for perfectly
rigid conducting elements ($\mu_1/\mu_0\rightarrow \infty$) which leads to
\begin{equation}
\label{GF7}
{\rm GF}=(2d/\xi)(1+s/d),
\end{equation}
so that the enhancement factor grows linearly with $s/d$.
 
The simplicity of Eq.(\ref{GF5}) stems from the assumption of 
one-dimensionality for which only one component of the stress and 
strain functions is needed. The analysis of a more realistic three
dimensional case is of course much more complex. In fact, in addition to
the complete solution of the strain field, one should also calculate the
actual current path which is given by hopping processes along directions
also different from the one where the field is applied.
However, as we show in the following section, some approximated analytical 
results can also be obtained for the three-dimensional case
by employing simplified models for the microstructure of TFRs.

\section{Three-dimensional case}
\label{3dim}

To analyze the piezoresistive response of a TFR let us consider
a cantilever beam in the $x$-$y$ plane with its
main axis lying on the $x$ direction. On top of the cantilever is deposited
the TFR with its thickness measured in the $z$ direction.
A bending of the cantilever beam produces a strain $\epsilon$ in 
the $x$ direction and, if the thickness of the cantilever is sufficiently small,
no strain in $y$. If $\nu$ is the Poisson ratio
of the cantilever, the strain along the $z$ direction is $-\epsilon\nu/(1-\nu)$.
The complete transfer of these strain values to the TFR leads to:
\begin{equation}
\label{strains1}
\bar{\epsilon}_{xx}=\epsilon, \,\,
\bar{\epsilon}_{yy}=0, \,\,
\bar{\epsilon}_{zz}=-\epsilon\nu/(1-\nu),
\end{equation}
where $\bar{\epsilon}_{xx}$, $\bar{\epsilon}_{yy}$, and $\bar{\epsilon}_{zz}$
are the macroscopic strain components within the TFR along the $x$, $y$, and $z$
directions. Within this setup,
the piezoresistive response is characterized by two distinct gauge factors:
the longitudinal gauge factor, ${\rm GF}_{L}$, obtained when the potential
difference is applied along the $x$ direction, and the transversal gauge factor,
${\rm GF}_{T}$, when the field is applied along the $y$ direction.
It is important to distinguish between intrinsic and geometric contributions
to the piezoresistive response. In fact, from quite general 
arguments,\cite{morten} it can be shown that, under the strain field of
Eq.(\ref{strains1}), ${\rm GF}_{L}$ and ${\rm GF}_{T}$ satisfy the following
relation:
\begin{equation}
\label{relation}
{\rm GF}_L-{\rm GF}_T=2+{\rm GF}^{\rm intr}_L-{\rm GF}^{\rm intr}_T,
\end{equation}
where the factor $2$ stems from the geometrical deformation of the
TFR, while ${\rm GF}^{\rm intr}_L$ and ${\rm GF}^{\rm intr}_T$ are intrinsic
gauge factors which are governed by microscopic transport properties
and are equal for isotropic systems. It is however experimentally
observed that ${\rm GF}_L-{\rm GF}_T>2$, suggesting that the
piezoresistive response of TFRs is affected by a certain amount of anisotropy.

\begin{figure}
\centerline{\psfig{figure=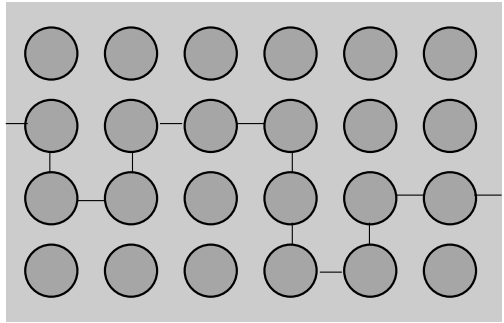,width=7cm}}
\caption{Simplified scheme of the transport pattern within the lattice
model. The spheres represent the conducting phase embedded in the
insulating medium (lighter region). Each sphere has diameter $s$
and the tunneling processes (indicated by the solid lines) take
place between the shortest sphere-to.sphere distance $d$.}
\label{fig1}
\end{figure}

To describe in a simple way the complex pattern of the current path and the
local strain distribution, we introduce a model in which a cubic lattice
of conducting spheres of diameter $s$ with near-neighbour distance $d$
is immersed in the insulating medium (Fig.1).
Moreover, the two phases have different isotropic elastic properties. 
As already assumed in the previous section, we disregard the contribution
to transport coming from the conducting phase and assume for the moment
direct tunneling between two neighbouring spheres. 
Within this arrangement of the conducting/insulating phases, we model the
geometry of the current path and the anisotropy by assuming that, in average,
charge transport can be described by
$N_{\parallel}$ tunneling processes parallel and
$N_{\perp}$ tunneling processes orthogonal to the direction of the
voltage drop.\cite{grima} Therefore, if the electric field is applied along the $x$ 
direction, the resulting tunneling resistance is $\widetilde{R}_x=N_{\parallel}R_x+
N_{\perp}(R_y+R_z)/2$, where $R_x$, $R_y$ and $R_z$ are resistances for
tunneling along the $x$, $y$, and $z$ directions, respectively.
We make the further approximation 
that $R_x$, $R_y$, and $R_z$ have the same average value. 
When the macroscopic strain field of Eq.(\ref{strains1}) is applied,
the different resistances respond to the corresponding local strains:
\begin{equation}
\label{strains2}
\frac{\delta R_x}{R_x}=\frac{2d}{\xi}\epsilon^0_{xx}, \,\,\,
\frac{\delta R_y}{R_y}=\frac{2d}{\xi}\epsilon^0_{yy}, \,\,\,
\frac{\delta R_z}{R_z}=\frac{2d}{\xi}\epsilon^0_{zz},
\end{equation}
where $\epsilon^0_{xx}=\delta_x d/d$, $\epsilon^0_{yy}=\delta_ y d/d$
and $\epsilon^0_{zz}=\delta_z d/d$. According to the previous definition
and to Eq.(\ref{strains2}), the longitudinal gauge factor reduces to:
\begin{eqnarray}
\label{strains3}
{\rm GF}^{\rm intr}_L\equiv \frac{\delta \widetilde{R}_x}{\bar{\epsilon}_{xx}
\widetilde{R}_x} & = & \frac{N_{\parallel}\delta R_x+N_{\perp}
(\delta R_y+\delta R_z)/2}
{\bar{\epsilon}_{xx}[N_{\parallel}R_x+N_{\perp}(R_x+R_y)/2]} \nonumber \\
& \simeq & \frac{2d}{\xi}\left[(1-\chi)\frac{\epsilon^0_{xx}}
{\bar{\epsilon}_{xx}}+\frac{\chi}{2}\left(\frac{\epsilon^0_{yy}}
{\bar{\epsilon}_{xx}}+\frac{\epsilon^0_{zz}}{\bar{\epsilon}_{xx}}\right)
\right]\, , \nonumber \\
\end{eqnarray}
where $\chi=N_{\perp}/(N_{\parallel}+N_{\perp})$.
By following the same lines, the intrinsic transversal gauge factor ${\rm GF}^{\rm intr}_T$
is estimated by using the resistance for a field applied along the
$y$ direction: $\widetilde{R}_y=N_{\parallel}R_y+N_{\perp}(R_x+R_z)/2$. Hence:
\begin{eqnarray}
\label{strains4}
{\rm GF}^{\rm intr}_T\equiv \frac{\delta \widetilde{R}_y}{\bar{\epsilon}_{xx}
\widetilde{R}_y} & = & \frac{N_{\parallel}\delta R_y+N_{\perp}
(\delta R_x+\delta R_z)/2}
{\bar{\epsilon}_{xx}[N_{\parallel}R_y+N_{\perp}(R_x+R_z)/2]} \nonumber \\
& \simeq & \frac{2d}{\xi}\left[(1-\chi)\frac{\epsilon^0_{yy}}
{\bar{\epsilon}_{xx}}+\frac{\chi}{2}\left(\frac{\epsilon^0_{xx}}
{\bar{\epsilon}_{xx}}+\frac{\epsilon^0_{zz}}{\bar{\epsilon}_{xx}}\right)
\right] . \nonumber \\
\end{eqnarray}
Note that piezoresistive anisotropy, 
${\rm GF}^{\rm intr}_L > {\rm GF}^{\rm intr}_T$,
is obtained for $\chi<2/3$.

At this point, the piezoresistive response can be calculated once the values
of the local strains are known. However, also for the simplified model
of spheres on a lattice, the general solution of the elastic problem is
difficult. On the other hand, analytical results can be obtained by 
employing some approximations aimed to simplify as much as possible the
elastic behavior of such a model. In the following we present two different
approximation schemes.

\subsection{Rigid spheres}
\label{limitb1b0}

In our model of transport described above, the current path is made
of tunneling processes between the shortest distances ($d$) between two
neighbouring spheres. Hence, the piezoresistive response is governed
by variations of $d$ along the $x$, $y$ and $z$ directions
due to the applied macroscopic strains of Eq.(\ref{strains1}).
These modified inter-sphere distances are easily found in the limiting
case of perfectly rigid spheres 
since in this case there is not deformation of the conducting 
phase ($\delta_xs=\delta_ys=\delta_zs=0$). In fact, if $L_x$,
$L_y$ and $L_z$ are the linear dimensions of the TFR, we have
\begin{equation}
\label{rigid1}
\bar{\epsilon}_{xx}\equiv \frac{\delta L_x}{L_x}=\frac{\delta_xd}{s+d},
\end{equation}
and
\begin{equation}
\label{rigid2}
\epsilon^0_{xx}\equiv \frac{\delta_x d}{d}=\bar{\epsilon}_{xx}(1+s/d),
\end{equation}
and the corresponding expressions for $\bar{\epsilon}_{yy}$ and 
$\bar{\epsilon}_{zz}$. Upon substitution in Eqs.(\ref{strains3},\ref{strains4}),
we find therefore:
\begin{equation}
\label{rigid3}
{\rm GF}^{\rm intr}_L  = \frac{2d}{\xi}\left(1-\frac{\chi}{2}
\frac{2-\nu}{1-\nu}\right)(1+s/d),
\end{equation}
and
\begin{equation}
\label{rigid4}
{\rm GF}^{\rm intr}_T  = \frac{2d}{\xi}\left(\frac{\chi}{2}
\frac{1-2\nu}{1-\nu}\right)(1+s/d).
\end{equation}
Although the tortuous character of the current path enters in the 
three-dimensional case (through the parameter $\chi$),
equations (\ref{rigid3}) and (\ref{rigid4}) have the same enhancement
factor $1+s/d$ as in the one-dimensional case
of Eq.(\ref{GF5}) in the limit $\mu_1/\mu_0\rightarrow \infty$.

\subsection{$s/d\gg 1$ limit}
\label{limitsd}

Another case which can be treated easily is given by assuming very
short distances $d$ compared to the size of the conducting spheres
($s/d\gg 1$). In this limit and in the region where tunneling takes place, 
the boundaries separating two neighbouring spheres can be treated
in  first approximation as planes parallel to each other. If we
neglect the shear components of the stress and strain fields, 
however expected to be weak in this region, we find that 
the stress and strain fields are locally constant within each phase
and equal to:\cite{ela}
\begin{equation}
\label{limitsd1}
\left(\begin{array}{c}
\sigma^i_{xx} \\
\sigma^i_{yy} \\
\sigma^i_{zz} \\ \end{array}\right)=
\left(\begin{array}{ccc}
\lambda_i+2\mu_i & \lambda_i & \lambda_i \\
\lambda_i & \lambda_i+2\mu_i & \lambda_i \\
\lambda_i & \lambda_i & \lambda_i+2\mu_i \\ \end{array}\right)
\left(\begin{array}{c}
\epsilon^i_{xx} \\
\epsilon^i_{yy} \\
\epsilon^i_{zz} \\ \end{array}\right) ,
\end{equation}
where $\lambda_i$ and $\mu_i$ are the Lam\'e coefficients
for the insulating ($i=0$) and conducting ($i=1$) phases.
Since at the boundaries between the two phases the stresses must
be equal, we find from Eq.(\ref{limitsd1})
\begin{equation}
\label{limitsd2}
\left(\begin{array}{c}
\epsilon^0_{xx} \\
\epsilon^0_{yy} \\
\epsilon^0_{zz} \\ \end{array}\right)=
\left(\begin{array}{ccc}
\alpha & \beta & \beta \\
\beta & \alpha & \beta \\
\beta & \beta & \alpha \\ \end{array}\right)
\left(\begin{array}{c}
\epsilon^1_{xx} \\
\epsilon^1_{yy} \\
\epsilon^1_{zz} \\ \end{array}\right) ,
\end{equation}
where
\begin{eqnarray}
\label{limitsd3}
\alpha & = &\frac{(\lambda_1+2\mu_1)(\lambda_0+\mu_0)-\lambda_1\lambda_0}
{(\lambda_0+2\mu_0)(\lambda_0+\mu_0)-\lambda_0^2} \nonumber \\
& = & \frac{B_1}{B_0}\left[1-2\frac{\nu_1-\nu_0}{(1+\nu_1)(1-2\nu_0)}\right],
\end{eqnarray}
and
\begin{eqnarray}
\label{limitsd4}
\beta & = & \frac{\mu_0\lambda_1-\mu_1\lambda_0}
{(\lambda_0+2\mu_0)(\lambda_0+\mu_0)-\lambda_0^2} \nonumber \\
& = & \frac{B_1}{B_0}\frac{\nu_1-\nu_0}{(1+\nu_1)(1-2\nu_0)}.
\end{eqnarray}
In the above expressions we have expressed the parameters $\alpha$ and $\beta$
in terms of the bulk modulus $B_i$ and the Poisson ratio $\nu_i$ of the
corresponding phase, which are linked to the Lam\'e coefficients by the relations:
\begin{equation}
\label{limitsd5}
\mu_i=\frac{3}{2}B_i\frac{1-2\nu_i}{1+\nu_i}; \,\,\,\,
\lambda_i=3B_i\frac{\nu_i}{1+\nu_i}.
\end{equation}
Note that if the spheres are almost touching, the applied strain
field (\ref{strains2}) is transfered mostly to the conducting phase and,
if we make the additional assumption that within the spheres the strains
are almost constant, we obtain
\begin{equation}
\label{limitsd6}
\left(\begin{array}{c}
\epsilon^1_{xx} \\
\epsilon^1_{yy} \\
\epsilon^1_{zz} \\ \end{array}\right)\simeq
\left(\begin{array}{c}
\bar{\epsilon}_{xx} \\
\bar{\epsilon}_{yy} \\
\bar{\epsilon}_{zz} \\ \end{array}\right) ,
\end{equation}
which is an approximate but indicative relation.
Finally, substituting Eqs.(\ref{limitsd1}-\ref{limitsd6}) into
Eqs.(\ref{strains3},\ref{strains4}), we find:
\begin{eqnarray}
\label{limitsd7}
{\rm GF}^{\rm intr}_L\simeq\frac{2d}{\xi}\Big[ & 1 & -\frac{\chi}{2}\frac{2-\nu}{1-\nu}
-\Big(1-\frac{3}{2}\chi\Big)\frac{2-\nu}{1-\nu} \nonumber \\
& \times &\frac{\nu_1-\nu_0}{(1+\nu_1)(1-2\nu_0)}\Big]\frac{B_1}{B_0}
\end{eqnarray}
for the longitudinal gauge factor and
\begin{eqnarray}
\label{limitsd8}
{\rm GF}^{\rm intr}_T\simeq\frac{2d}{\xi}\Big[&\frac{\displaystyle \chi}{\displaystyle 2}&
\frac{1-2\nu}{1-\nu}
+\Big(1-\frac{3}{2}\chi\Big)\frac{1-2\nu}{1-\nu} \nonumber \\
&\times&\frac{\nu_1-\nu_0}{(1+\nu_1)(1-2\nu_0)}\Big]\frac{B_1}{B_0}
\end{eqnarray}
for the transversal one.

The main result of Eqs.(\ref{limitsd7},\ref{limitsd8}) is that the
gauge factors are proportional to the ratio of the bulk moduli 
$B_1/B_0$ which, as long as $B_1 >B_0$, determines the enhancement
of the piezoresistive response with respect to the homogeneous
limit $B_1=B_0$. In RuO$_2$-based TFRs, this enhancement factor
can in principle be as large as $B_1/B_0\simeq 4-5$.
Note however that hardening of the glassy phase due to
diffusion of Ru particles ($B_{\rm Ru}\simeq 220$ GPa) could 
reduce $B_1/B_0$ lowering therefore the piezoresistive response.

\begin{figure}
\centerline{\psfig{figure=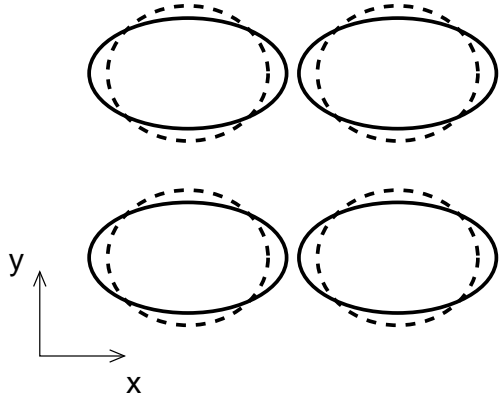,width=7cm}}
\caption{Pictorial explanation of the effect of different
Poisson's ratios of the conducting ($\nu_1$) and insulating
($\nu_0$) phases. It is assumed that the macroscopic strain
is only along the $x$ direction. The dashed and solid lines 
represent the contours of the elastically deformed conducting
spheres when $\nu_1=\nu_0$ and $\nu_1 > \nu_0$, respectively.
For an electric field applied along the $x$ direction. the gauge 
factor is diminished while for a field along the $y$ direction
the transversal response is enhanced.}
\label{fig2}
\end{figure}

Another interesting feature is the effect of the
Poisson's ratios of the conducting ($\nu_1$) and insulating ($\nu_0$)
phases. In fact, since we consider $\chi < 2/3$, when $\nu_1 > \nu_0$ 
the longitudinal (transversal)
gauge factor is diminished (enhanced) with respect to the limit
$\nu_1=\nu_0$. For $\nu_1 < \nu_0$ this situation is reversed.
This is explained by noting that Poisson's ratio measures the
lateral contraction  relative to the longitudinal extension
for a homogeneous material in a tensile stress. Therefore,
in the heterogeneous system we consider, the phase with higher
Poisson's ratio will have a more pronounced transversal contraction
than the phase with lower value of $\nu$. This affects the piezoresistive
response as schematically illustrated in Fig. 2, where a cut along
the $x$-$y$ plane of our model of TFRs is shown.
For simplicity, we consider that the only nonzero macroscopic strain 
component is $\bar{\epsilon}_{xx}=\epsilon$ [{\it i.e.}, $\nu=0$
in Eq.(\ref{strains1})]. The dashed and solid lines represent the
contour of the spheres deformed by $\bar{\epsilon}_{xx}$ when
$\nu_1=\nu_0$ and $\nu_1>\nu_0$, respectively. Compared to the
$\nu_1=\nu_0$ case, when $\nu_1>\nu_0$
the inter-sphere distances are reduced in the $x$ direction and
enhanced in the $y$ direction. Therefore, in the limiting case in which all the
tunneling processes are aligned with the voltage drop ($\chi=0$) we
expect that the longitudinal GF is reduced while the transversal
one is enhanced, in agreement with the trend given by 
Eqs.(\ref{limitsd7},\ref{limitsd8}). Of course, finite values of $\chi$ 
tend to lower this effect which vanishes in the isotropic limit $\chi=2/3$.

\section{Discussion and Conclusions}
\label{concl}
The results presented above clearly indicate two essential elements
which favour high piezoresistive responses in TFRs: \\
(i) the conducting phase must be stiffer than the glassy one ($B_1 > B_0$), \\
(ii) the mean linear size $s$ of the metallic particles
must be much larger than the typical tunneling distance $d$.

Both points (i) and (ii) seem to justify some of the trends and
properties of TFRs. For example, the use of hard oxide-metals as
constituent of the metallic phase is in agreement with (i),
while the observed enhancement of the gauge factor with the mean
metallic grain size\cite{carcia,prude2,prude3} is in conformity with (ii).
Hence, the problem of how the fabrication processes affect the piezoresistive
response could be approached by investigating the effects on (i)
and (ii). For example, the reduction of ${\rm GF}$ at high
peak firing temperature $T_{\rm f}$ could be due to
an enhanced dispersion of the metallic phase into the glass. This would
enhance the effective bulk modulus of the glass leading to a lower
ratio $B_1/B_0$.

In conclusion, we have proposed a theory of piezoresistivity in TFRs
based on elastic heterogeneity. Within a simplified model of TFRs, we
have shown that when the conducting phase is stiffer than the insulating
one, the local strains within this latter are enhanced with respect
to the averaged macroscopic strains. This enhancement leads to
higher piezoresistive responses leading to higher GF values
compared to those expected for a homogeneous system.

\acknowledgments
The authors would like to thank M. Prudenziati
for interesting discussions.



\begin{references}
\bibitem{note1}
Here we disregard additional temperature dependences given by
thermal expansion, intra-grain electron-phonon scattering,
activation energies, etc.. These effects can be added,
however they are not important for order-of-magnitude
arguments presented here.
\end{references}
\end{document}